\documentclass[prb,twocolumn,showpacs,superscriptaddress,floatfix, nofootinbib]{revtex4-2}
\usepackage{amsmath,amssymb,amsfonts,mathrsfs, amsbsy}
\usepackage{graphicx}
\usepackage{lipsum, babel}
\usepackage[bookmarks=true, colorlinks=true, linkcolor=blue, urlcolor=blue, citecolor=blue, bookmarks=true, hyperindex=true]{hyperref}
\usepackage{comment}

\newcommand{\be}{\begin{equation}}
\newcommand{\ee}{\end{equation}}

\newcommand{\beq}{\begin{eqnarray}}
\newcommand{\eeq}{\end{eqnarray}}

\def\H1{\widehat{H}_1}

\begin{document}

\title{Tomonaga-Luttinger liquid-Bose glass phase transition in a system of 1D disordered fermions with pair hoppings}

\author{M.\,S.~Bahovadinov}
\affiliation{ Physics Department, National Research University Higher School of Economics, Moscow 101000, Russia}
\affiliation{Russian Quantum Center, Skolkovo, Moscow 143025, Russia}
 \author{R.\,O.~Sharipov}
 \affiliation{ Physics Department, Faculty of Mathematics and Physics, University of Ljubljana,
Ljubljana, Slovenia}
 \affiliation{Russian Quantum Center, Skolkovo, Moscow 143025, Russia}
  \author{B.\,L.~Altshuler}
  \affiliation{Physics Department, Columbia University, 538 West 120th Street, New York, New York 10027, USA}
  \affiliation{Russian Quantum Center, Skolkovo, Moscow 143025, Russia}
  \author{G.\,V.~Shlyapnikov}
\affiliation{Russian Quantum Center, Skolkovo, Moscow 143025, Russia}
\affiliation{Moscow Institute of Physics and Technology, Dolgoprudny, Moscow Region, 141701, Russia}
\affiliation{Université Paris-Saclay, CNRS, LPTMS, 91405 Orsay, France}
\affiliation{Van der Waals–Zeeman Institute, Institute of Physics, University of Amsterdam, Science Park 904, 1098 XH Amsterdam, The Netherlands}
\date{\today}

\begin{abstract}
We consider a one-dimensional system of spinless fermions with single particle and pair hoppings  in a potential on-site disorder. The pair hopping term violates integrability of the model and enhances superconducting fluctuations in the bulk of the liquid. By means of the Abelian bosonization and extensive numerical DMRG calculations we provide evidence that sufficiently large pair hopping amplitude guarantees the survival of the Tomonaga-Luttinger liquid phase at weak disorder. Large disorder drives the system to the Bose glass phase, realising the Giamarchi-Schulz scenario in such a system. 
\end{abstract}

\maketitle

\section{Introduction}  
An interplay between interparticle interactions and disorder in low-dimensional quantum many-body systems has been extensively studied during the last decades~\cite{AronovAltshuler,Apel, Giamarchi, Altshuler} and currently remains an active research frontier (for reviews, see e.g. Refs~\cite{ ReviewLuitz, ReviewAbanin}). At high energies many studies of interacting quantum many-body systems support a strong indication of a transition to the many-body localized (MBL) phase at sufficiently strong disorder. In the MBL phase the eigenstate thermalization hypothesis (ETH) is violated~\cite{Pal_Huse_2010, YBLev2014, Serbyn2015, Luitz2016, De_Luca_2013, De_Luca_2014}, which leads to the protection of quantum states from decoherence. 

\begin{figure}[t]
\includegraphics[width=\columnwidth]{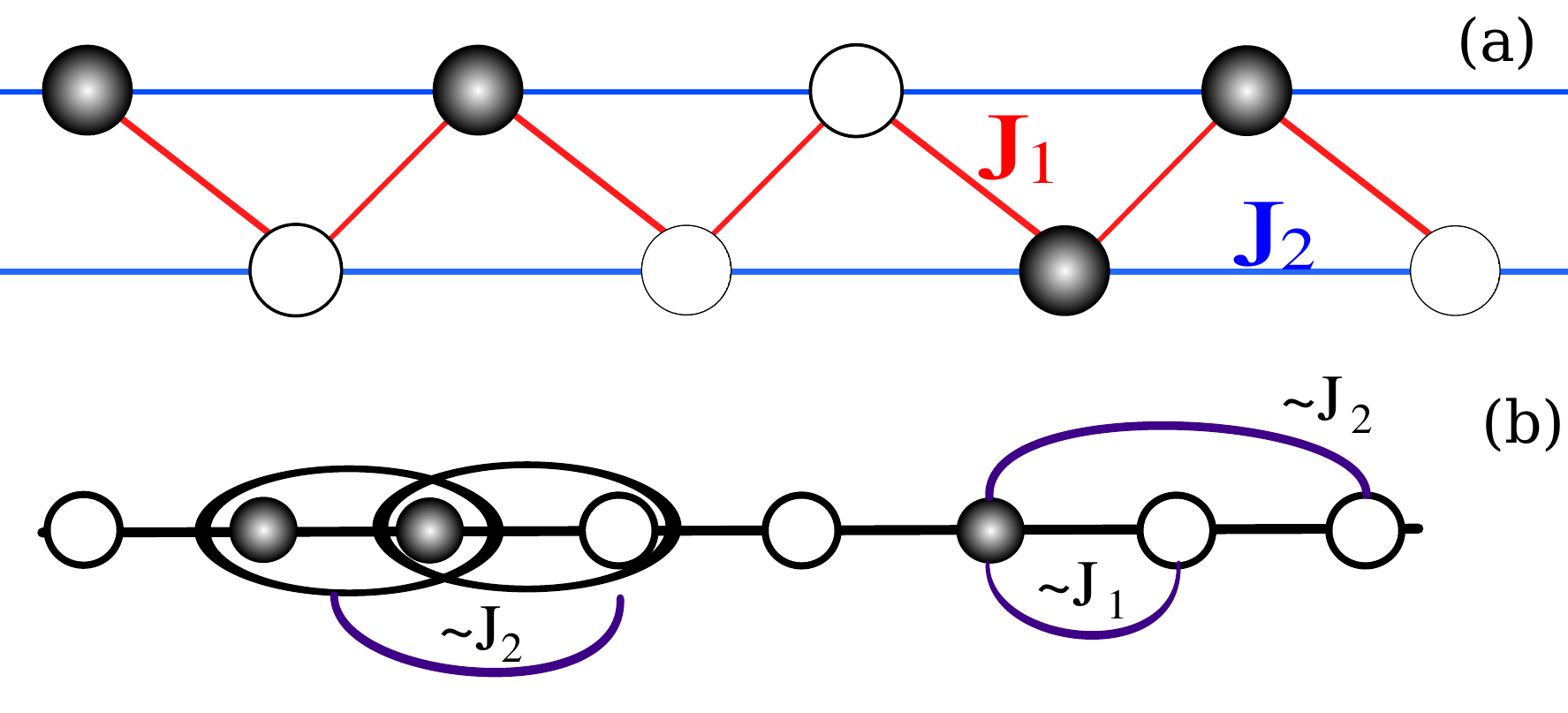}
\caption{(a) Schematic representation of the fermionic model (\ref{H}). For $J_2$/$|J_1|<0$ on top of single particle hoppings pair hopping events are also favoured, as sketched in (b).}
\label{fig:r}
\end{figure}
Although studies of the MBL phenomenon are currently under active research, an interplay between disorder and interactions in the low-energy sector is still under investigation, especially in one-dimensional quantum systems, where quantum fluctuations are the strongest. In the clean limit in 1D, gapless phases are traditionally described within the Tomonaga-Luttinger liquid (TLL) theory with algebraically decaying correlations at large distances, controlled by the TLL parameter $K_0$. It is natural to expect that arbitrarily weak disorder drives the TLL to Bose glass (BG) phase. 
However, in their seminal paper based on perturbative treatment of disorder Giamarchi and Schulz (GS)~\cite{GS1,GS2} have shown that for the TLL with the clean parameter $K_0 > 3/2$ the Berezinskii-Kosterlitz-Thouless (BKT) quantum phase transition to BG phase should occur for a small but finite disorder strength. Two-loop calculations confirmed this picture, with a jump of $K$ to $0$ at the BKT transition occuring at the critical value $K=K_c = 3/2$~\cite{R1,R2}. For TLL with $K_0<3/2$, superfluid (SF) phase is destroyed in an arbitrarily weak disorder with the resulting correlation length $L_c= W^{-2\phi_s}$, where $W$ is the disorder strength and $\phi_s=(3-2K_0)^{-1}$. For strong disorder, alternative scenarios with $K_c>3/2$ were previously proposed~\cite{Altman1, Altman2, Altman3,Boris,Pollet2013,Pielawa2013,Pollet2014,Yao}. We further refer to them as weak link scenarios. The recent numerical work\cite{Doggen2017} demonstrated the presence of such weak-link scenario at strong-disorder criticalities for the disordered 1D XXZ model.

The traditional recipe to observe SF-BG transition in 1D spin-1/2 quantum magnets is to include strong ferromagnetic (FM) Ising interaction~\cite{DotyFisher,RungeZimanyi,Eckern,Urba,Poboiko,BoseGlass1,BoseGlass2,Doggen2017} (this term fermionizes to the attraction between two neighboring fermions). In this case, pairing fluctuations in the bulk can be strongly enhanced with resulting $K_0>3/2$, so that sufficiently weak disorder is unable to localize the ground state. On the other hand, strong Ising interactions are known to cause phase transition to the ferromagnetic phase, limiting the region where the SF-BG phase transition can be observed. For the 1D XXZ model this region is bounded with $-1<\Delta<-1/2$~\cite{DotyFisher,RungeZimanyi,Eckern,Urba}, where $\Delta$ is an amplitude of the Ising interaction.

Enhanced superconducting correlations in the system of spinless fermions have been studied in several lattice models in 1D~\cite{Mattioli2013,Dalmonte2015,Kane2017,He2019}. A standard way to introduce pairing correlations is through density-density interactions as in the XXZ magnet case. Alternatively, one can enhance these correlations via pair hopping terms. Recently, a model with this feature was studied in 1D by J. Ruhman and E. Altman~\cite{RuhmanAltman}. Although the Ruhman-Altman model is rather abstract, it got sufficient attention and the phase diagram of this model was recently studied numerically by means of the Density Matrix Renormalization Group (DMRG) method~\cite{Mazza,Mazza2}.

In this work we study the model of 1D spinless fermions with single particle and pair hoppings in a random potential, which is dual to the model of hard-core bosons (maximum occupation is 1 boson per lattice site) with the nearest-neighbor and the next-nearest-neighbor hoppings. 
The model of our study can be experimentally realized in several systems (using 3D transmon qubits on saphire and also using ultracold bosonic atoms in optical lattices), as suggested by recent proposals~\cite{Proposal1,Proposal2}. The presence of the pair hopping term breaks integrability of the model and shifts the TLL parameter $K_0$ from unity.

In the clean limit, we first construct an effective low-energy field theory via Abelian bosonization of the model, compare analytical expressions with numerical results and obtain the phase diagram at $T=0$. Following the GS method we show that if pairing correlations are sufficiently strong, then the SF phase survives in a weak disorder, whereas strong disorder drives the system to the BG phase. We provide numerical confirmation of these statements. Importantly, in contrast to the weak-versus-strong disorder scenarios reported for the disordered 1D XXZ (spin-1/2) model~\cite{Doggen2017}, in our model we observe only the GS scenario with $K_c=3/2$ for the considered parameter space. Our numerical results are based on the DMRG method in its tensor-network formulation (with conserved U(1) symmetry)~\cite{White1,White2,Schollwock,Rizzi}.       

The paper is organized as follows. In Sec.~\ref{S_model_spin} we present our fermionic model. To characterize the phase diagram of the model in the clean and disordered cases we use a set of quantities introduced in section Sec.~\ref{Probes}. In Sec.~\ref{Bosonization} we present the low-energy TLL theory of the clean model obtained by a constructive bosonization procedure.  Our numerical results are demonstrated in Sec.~\ref{ResClean} for the clean case and in Sec.~\ref{ResDis} for the disordered case. Sec.~\ref{Conc} is devoted to our concluding remarks.

\section{Model and symmetries} 
\label{S_model_spin}

We consider a 1D system of fermions with the single-particle and pair hopping terms in the zig-zag ladder of (even) $L$ sites with the periodic boundary condition (see Fig.~(\ref{fig:r})). The Hamiltonian

\be \label{H}
	H =  H_1+H_2+H_{dis},
\ee
contains the single hopping $H_1$ and pair hopping $H_2$ terms:
\be \label{H_1}
	H_1 =  \sum_{\beta=1,2} \sum_{j=1}^{L} \frac{J_\beta(-1)^{\beta+1}}{2} \left(c^\dagger_i c_{i+\beta} + h.c. \right), 
\ee
\be \label{H_2}
H_2=J_2\sum_i^{L} \left( c^\dagger_{i}c_{i+1} c^\dagger_{i+1}c_{i+2} +h.c. \right),
\ee
where $J_1$ and $J_2$ are the single particle and pair hopping amplitudes, and we impose $c_{L+1}=c_1$.
As the disorder term we consider a random onsite potential
\be \label{H_dis}
H_{dis} = \sum_i h_i (\hat{n}_i-1/2).
\ee
The on-site potential realizations $h_i$ are drawn from the uniform distribution $h_i \in [-h,h]$. We further refer to $h$ as a disorder strength/amplitude. For convenience we also introduce a parameter $\kappa=\frac{J_2}{|J_1|}$ and consider $N/L$=1/2, where $N$ is the total number of fermions in the ladder (half-filling).

In the presence of disorder Hamiltonian (\ref{H}) conserves only the particle number (U(1) symmetry). If $\kappa=0$ or $\kappa=\infty$, the eigenstates are localized for an arbitrary $h$, since one can map the model exactly onto the model of free disodered particles in 1D. Thus, at these values of $\kappa$ the model (\ref{H}) is integrable in both the disordered and clean cases.
For $\kappa \ll 1 $ the clean model is quasi-integrable, possessing
quasi-conserved charges, as was shown in the recent works~\cite{Kurlov2021,Bahovadinov2022,Kurlov2023,Motrunich2023}. We consider only $-1< \kappa \lesssim 0.33$, where, as shown below, the TLL theory serves as a valid framework of $T=0$ physics. We emphasize that Hamiltonian (\ref{H}) can be mapped exactly via Jordan-Wigner transformation~\cite{JWT} onto the XY model ($s=1/2$) in the zig-zag ladder in a random magnetic field~\cite{Bahovadinov2022}:
\be \label{H_spin}
 H =  \sum_{\beta=1,2}J_\beta \sum_{i=1}^{L} \left[ S^x_i S^x_{i+\beta}+S^y_i S^y_{i+\beta} \right]+ \sum_{i=1}^L h_i S_i^z.
 \ee
 Recently, it was shown in Ref.~\cite{Bahovadinov2022} that this model exhibits the MBL transition guaranteed by the pair-hopping term (\ref{H_2}). 

At $T=0$ one expects that for $\kappa<0$ the SF phase is conserved with the modified TLL parameter $K_0>1$. Indeed, pair-hopping of fermions amplifies SF correlations, resulting in a slower algebraic decay of these correlations. On the contrary, for $\kappa>0$, one has dominating charge-density wave type correlations, with $K_0<1$. To highlight this, one can rewrite Eq.~(\ref{H_2}) as a correlated hopping term $-\kappa(c^\dagger_in_{i+1}c_{i+2} +h.c.)$: hopping of fermions along a given leg pins fermionic density on the other leg. If $\kappa>0$ and is sufficiently large, then such pinning can cause spontaneous dimerization onto the $2k_F$ bond-order density wave~\cite{Haldane1982}. This phase transition was previously shown~\cite{Haldane1982,Okamoto1993,Hirata1999, Lecheminant2001,Sota2010,Mishra2013} to occur in the vicinity of the critical $\kappa$, which in the clean case is $\kappa_c \approx 0.33$. In the following sections, we first estimate the critical point $\kappa_c$ using bosonization and then use the DMRG method to locate the transition point with an improved accuracy.
\section{Numerical method and calculated quantities}
\label{Probes}
 In this section we present details of our numerical calculations and introduce necessary quantities to characterize phases of the clean and disordered models. For numerical convenience we considered the model in its spin-1/2 representation (\ref{H_spin}).
    We used variational two-site DMRG algorithm to obtain accurate matrix product state representation of the ground state in the half-filled sector of the Hilbert space from a given product state. Due to the variational character of the algorithm, the convergence to the ground state is not guaranteed, especially in the disordered case. Thus, we introduce a cascade of noise during the sweeping procedure with a vanishing amplitude in each step of the cascade. We used a large number of sweeps and gradually increased bond-dimensions up to $\chi \approx 3000$ for the largest considered system sizes in the clean case. For the disordered case this value has reached $\chi \approx 1200$. These steps guarantee the convergence to the ground state during the RG procedure. Largest truncation errors in the last sweepings were of the order of $\epsilon\sim 10^{-10}$.  
For the purpose of the work we use the following quantities to characterize phases of the clean and disordered models.

{\it{Central charge in the clean case.} }
 Our model at $\kappa=0$ is critical and belongs to the (1+1) dimensional Gaussian universality class with the central charge $c=1$\cite{GogolinBook}. The latter quantity is kept fixed in the TLL phase, whereas for the gapped (insulating) phase we have $c=0$. We calculate the central charge using the expression for the von Neumann entanglement entropy (EE). EE of the subsystem with length $l$ is defined as
$S_L(l)=-{\rm Tr}_l \rho_l \log \rho_l$, where $\rho_l={\rm Tr}_{L-l}\rho$ is 
the reduced density matrix of the subsystem and $\rho$ is the full density matrix of the whole system with length $L$. Using the Conformal Field Theory (CFT), EE of the subsystem with length $l$ was derived as:~\cite{Affleck91,Holzhey94,Calabrese04}
\begin{eqnarray}
S_L(l)=\frac{c}{3}\ln\left[\frac{L}{\pi}\sin\left(\frac{\pi l}{L}\right)\right]+b,
\label{EE}
\end{eqnarray}
where the prefactor $c$ is the central charge and $b$ is a non-universal constant. This  expression is useful for the estimation of the central charge and, hence, it provides information on the universality class of the underlying CFT. From Eq.(\ref{EE}) one can easily obtain the following expression for the central charge~\cite{Nishimoto}: 
\begin{eqnarray}
c(L)=\frac{3\left[S_L\left(\frac{L}{2}-1\right)-S_L\left(\frac{L}{2}\right)\right]}
{\ln\left[\cos\left(\frac{\pi}{L}\right)\right]}.
\label{c}
\end{eqnarray}
Using this formula we can obtain the central charge for a 1D system with a fixed length $L$ in a ring geometry. We use the calculated central charge to show the transition to the gapped bond-order wave phase for $\kappa_c\approx 0.33$.
 
{\it{TLL parameter $K_0$ in the clean case.} }
Although the central charge correctly captures the transition to the gapped phase, we exploit bipartite fluctuations of magnetization to evaluate the critical $\kappa_c$ with an improved accuracy. This quantity was shown to be an efficient probe to capture quantum critical points in low-dimensional quantum systems~\cite{Nishimoto,Bipartite,Bipartite2}. It is related to the magnetization fluctuation of subsystem $A$ with length $l$, 
${\cal{F}}_L(l)=\langle(\sum_i S^z_i-\sum_i \bar{S}^z_i)^2\rangle$, where $i$ belongs to the subsystem $A$ with the average magnetization $\sum_i \bar{S}^z_i$, and the fluctuation behaves as~\cite{Song2010}
\begin{eqnarray}
 {\cal{F}}_L(l)=\frac{K_0}{\pi^2} \ln\left[\frac{L}{\pi}\sin\left(\frac{\pi l}{L}\right)\right]-\frac{(-1)^lb_0}{\left[\frac{L}{\pi}\sin\left(\frac{\pi l}{L}\right)\right]^{2K_0 }}+b_1,
\label{Fluctuation}
\end{eqnarray}
where $b_0$ and $b_1$ are non-universal constants. Bipartite fluctuations ${\cal{F}}_L(l)$ behave similarly to $S_L(l)$.
Thus, one obtains an expression similar to  Eq.(\ref{c})~\cite{Nishimoto}:
\begin{eqnarray}
 K_0(L)=\frac{\pi^2\left({\cal{F}}_L\left(\frac{L}{2}-2\right)-{\cal{F}}_L\left(\frac{L}{2}\right)\right)}{\ln\left[\cos\left(\frac{2\pi}{L}\right)\right]}.
\label{K_L}
\end{eqnarray}
In the derivation of Eq.(\ref{K_L}) we took into account the fact that for $\kappa>0$  the ${\cal O}(L^{-2K_0})$ correction given by the second term in Eq.(\ref{Fluctuation}) oscillates on alternating sites. Thus, ${\cal{F}}_L(\frac{L}{2})$ and ${\cal{F}}(\frac{L}{2}-2)$ is a more relevant choice. The reason of using this formula is to obtain an accurate estimnation of the TLL parameter $K_0$ within the parameter space $-1<\kappa \leq 0.33$. It also allows one to locate the critical value $\kappa_c$, where  $K_0=\frac{1}{2}$.

{\it{ TLL parameter $K$ in the disordered case.} }
For an accurate estimation of the TLL parameter $K$
for the disordered model, we calculate the single-particle density matrix $G_{i,j}=\langle S^+_iS^-_j \rangle$ and then extract the value of $K$ from the expected algebraic decay $\propto r^{-1/2K}$ in the SF phase, where $r=|i-j|$. To take into account finite-size effects in the periodic boundary condition setting, one replaces $r$ with an effective $\Tilde{r}=crd(r)$, where the chord function is defined as $crd(r)=\frac{L}{\pi}\sin(\pi r/L)$. Then, using the fitting $\log(C)=-\frac{1}{2K}\zeta+const.,$ with $C(r)=\frac{1}{L}\sum_j \langle S^+_jS^-_{j+r}\rangle$ and $\zeta=\log(\Tilde{r})$, we extract the value of $K$. In the GS case $K_c=3/2$, whereas in the {\it weak-link scenario} studied in Ref.~\cite{Doggen2017} one has $K_c>3/2$. To confirm the GS scenario in the whole parameter space with $K_c \approx 3/2$, we assume that $\log(C)=\frac{1}{2K}\zeta-\alpha\zeta^2+const.,$ with $\alpha>0$ in the BG phase. The latter is guaranteed due to the exponential decay of correlations and one can similarly extract the value of $\alpha$ from the fitting procedure. The increase of $\alpha$ with $h$ turns out to be sharp in the vicinity of the transition with $K_c=3/2$. This method was successfully exploited to capture the SF-BG transition in the disordered 1D Bose-Hubbard model~\cite{BoseHubbard}.  
\section{Bosonization and GS procedure}
\label{Bosonization}
We follow a constructive bosonization procedure to achieve an effective low-energy theory of the clean model. For this, the clean Hamiltonian (\ref{H}) is rewritten in the $k$-space:
\be\label{Hk}
\Tilde{H}=\sum_{k\in BZ} \epsilon_k c^\dagger_k c_k + \frac{J_2}{L}\sum_{k_1,k_2,q} \cos(2k_1+q)c^\dagger_{k_1+q}c_{k_1}c^\dagger_{k_2-q}c_{k_{2}},
\ee
with the single-particle dispersion relation:
\be
\epsilon_k=-\sum_{\beta=1,2} J_\beta\cos(\beta k).
\ee
In the weak-coupling regime, $|\kappa|\ll1$,  one starts with a linearized spectrum of the free fermionic term (\ref{H_1}) with the corresponding left (L) and right (R) moving branches. The first term of Eq.~(\ref{Hk}) can be rewritten as, 
\be \label{H0RHO}
\Tilde{H_0}=\frac{\pi v_F}{L} \sum_{q,\tau} \hat{\rho_{\tau}}(q)\hat{\rho_{\tau}}(-q),  
\ee
with $\tau \in [ L(-1), R(+1) ]$, where the Fermi velocity is $v_F=\frac{\partial\epsilon_k}{\partial k}|_{k=k_F}$ and the density plasmons for a given branch $\tau\in L,R$ are defined as,
\begin{equation} 
\hat{\rho}_\tau(q)=\sum_k c^\dagger_{\tau,k+q} c_{\tau, k}.
\end{equation}
Canonical fermionic operators $c^{(\dagger)}_{k,\tau}$ correspond to the $\tau$ branch. The second term of Eq.~(\ref{Hk}) can not be directly expressed in terms of these plasmons due to the $k$-dependence of the amplitude $V(k,q)=\cos(2k+q)$. However, for $|\kappa| \ll 1$  one can assume that  $V(k,q)\approx V(q,k_F)$, since the momentum of excitations $q$ is close to zero for the forward scattering, and $q\sim 2k_F$ for the back-scattering processes. One is left with the $k_F$ dependence of the scattering amplitudes $V(q\sim 0)=\cos(2k_F)$ and $V(q\sim 2k_F)=\cos(4k_F)$. This is expected, since if the density of particles (holes) exceeds half-filling, pair-hopping events are less probable and the effects of the corresponding term are weak, i.e the largest contribution is expected at half-filling. Within this approximation, one can rewrite the second term of Eq.~(\ref{Hk}) in terms of plasmonic excitations and fully bosonize the fermionic theory, since $[\hat{\rho}_\tau(-q),\hat{\rho}_{\tau^\prime}(q^\prime)]=\frac{Lq\tau}{2\pi}\delta_{\tau,\tau^\prime}\delta_{q,q^\prime}$. 
One then follows the standard bosonization scheme~\cite{GogolinBook, GiamarchiBook, Maslov} by introducing the conjugated bosonic fields
\be
\phi(x)=i\sum_{q\neq0} \frac{sgn(x)}{\sqrt{2|q|L}}(b^\dagger_qe^{-iqx}-b_qe^{iqx}),
\ee
and
\be
\theta(x)=-i\sum_{q\neq0} \frac{1}{\sqrt{2|q|L}}(b^\dagger_qe^{-iqx}-b_qe^{iqx}),
\ee
with $[\phi(x),\partial_{x^\prime}\theta(x^\prime)]=i\delta(x-x^\prime)$. As a result, (1+1) dimensional Sine-Gordon model is obtained:
\be
\label{SG}
H=\frac{v}{2}\int \left( \frac{1}{K_0}(\partial_x\phi)^2+K_0(\partial_x\theta)^2 \right) + g \cos(\beta_s\phi),
\ee
where $\beta_s=\sqrt{16\pi}$ and $v$ is the excitation velocity. The cosine term in Eq.~(\ref{SG}) originates from the $4k_F$ umklapp scattering, since we consider the half-filled sector of the Hilbert space. Expressions for the TLL parameter $K_0(\kappa,k_F)$ and the excitation velocity $v(\kappa,k_F)$ have the following forms:
\be
K_0(\kappa,k_F)=\sqrt{\frac{2\pi+4\kappa\left(3\cos(2k_F)-1\right)}{2\pi+4\kappa\left(\cos(2k_F)+1\right)}},
\ee
and 

\be
v(\kappa,k_F)=v_F\sqrt{\left(1+\frac{4\kappa\cos(2k_F)}{\pi}\right)^2-\left(\frac{4\kappa \sin(k_F)}{\pi}\right)^2  }.
\ee 
At half-filling ($k_F=\frac{\pi}{2}$) the expression for $K_0$ transforms to
\be
\label{K_expression}
K_0=\sqrt{1-\frac{8\kappa}{\pi}}.
\ee
 
For $K_0<1/2$ the cosine term in Eq.(\ref{SG}) becomes relevant in the RG sense and opens a gap in the spectrum via the BKT transition. From Eq.(\ref{K_expression}) we find the critical value  $\kappa_c=\frac{3\pi}{32} \approx 0.295.$ As we show in the next section, the found estimate of $\kappa_c$ is in good agreement with the numerical DMRG result.
 \begin{figure}[t]
\includegraphics[width=\columnwidth]{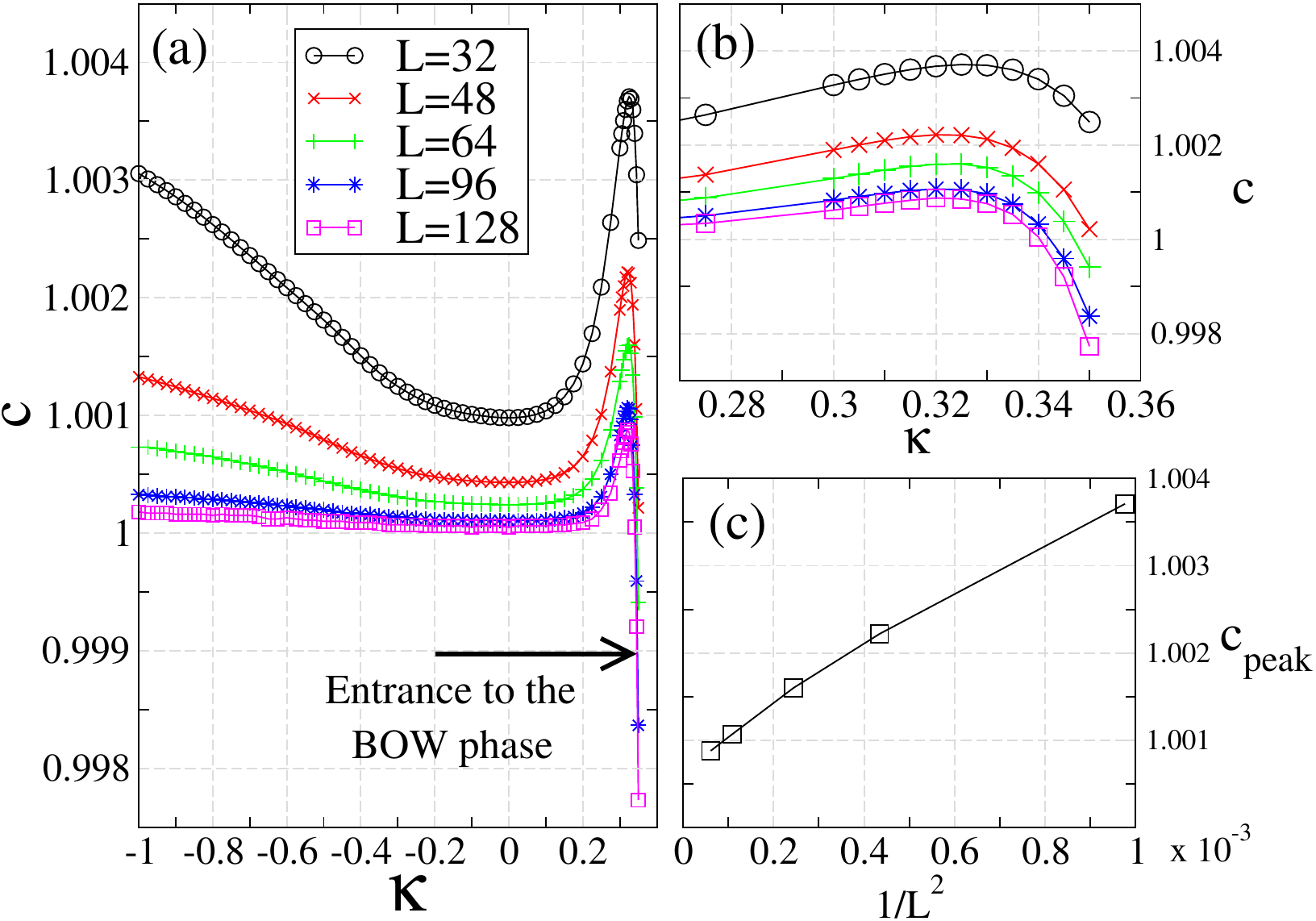}
\caption{ (a) Numerical DMRG results for the central charge $c$ versus $\kappa$. The results are obtained using Eq.~(\ref{c}) for the system sizes $L= \lbrace 32,48,64,96,128 \rbrace$. The pronounced peak values of $c$ are represented in the subplot (b). In (c) the scaling of the peak values $c_{peak}$ as $1/L^2$ is plotted. }
\label{fig:C}
\end{figure}

From Eq.~(\ref{SG}) it is clear that in the gapless regime all theory reduces to the theory of disordered TLL with the TLL parameters $v$ and $K_0$ given above. For weak disorder, we follow the RG formulation of GS~\cite{Giamarchi} and below in this section shortly present the well-known results. Assuming weak disorder, one rewrites the disordered part of the Hamiltonian as,
\begin{equation}
    \label{HdisC}
    H_{dis}=-\int dx \left[ ~\eta(x) \frac{1}{\sqrt{\pi} } \partial_x \phi+ \left( \xi(x)\frac{e^{-i \sqrt{4 \pi } \phi}}{2 \pi a}  +h.c. \right) \right],
\end{equation}
where the introduced Gaussian complex fields  $\eta(x)$  and $\xi(x)$  correspond to the scattering with momenta $q\sim0$ and $q\sim2 k_F$, respectively. Forward scattering is irrelevant
within the considered approximations, since
the corresponding term in the Hamiltonian can be
eliminated completely by the redefinition of the phase field $\phi(x)$. Then, one is left with the backscattering part of Eq.~(\ref{HdisC}) with the momentum transfer $q\sim2k_F$ and  $\overline{\xi(x) \xi^{*}(x')}=D~ \delta(x-x')$.
The disorder constant is $D=\langle h^2\rangle$ and the TLL constants $v$ and $K$ change under the RG procedure. Following the RG formulation of GS~\cite{Giamarchi}, one gets the following set of RG equations:
\begin{subequations}
\begin{equation}
   \frac{d\tilde{D}}{d l}= (3-2 K) \tilde{D},
\end{equation}
\begin{equation}
\frac{dK}{dl}=-\frac{ K^2}{2 } \tilde{D},~~~~\frac{dv}{dl}=-\frac{v K}{2} \tilde{D},
\end{equation}
\end{subequations}
where $l$ is the scaling RG parameter and $\tilde{D}=D \frac{2 a}{\pi v^2}$.  At finite disorder there is a critical value $K_h$, such that for $K<K_h$ the disorder flows to the strong-coupling localized phase, whereas for $K \geq K_h$ the disorder flows to zero under the RG transformation renormalizing  the bare parameters $v$ and $K$. The critical value of the TLL parameter depends on the disorder, but in the small disorder limit one has $K_h(h\rightarrow 0)=K_c=3/2$.
\section{Numerical results: clean case}
\label{ResClean}
{\it{Central charge.}}  Before considering the disordered case, we show our results for the clean case. We first demonstrate the results for the central charge in the parameter space $-1<\kappa<0.35$, where the upper bound is chosen to be close to the expected critical value $\kappa_c=0.33$. For this purpose, we performed DMRG calculations with the maximum bond-dimensions up to $\chi=3000$ and for the system sizes $L=\lbrace 32,64,96,128 \rbrace$. This allowed us to calculate the central charge up to five digits using Eq.~(\ref{c}). Our results are shown in Fig.~\ref{fig:C}. 
    The obtained values of $c$ for all system sizes are close to unity, which signals Gaussian universality class of the current (1+1) dimensional system. In this case the SF phase is stable. Important peculiarity of the presented plot is that for all system sizes one clearly observes a sharp peak of $c$ with an abrupt decrease in the vicinity of $\kappa\approx 0.33 $. This feature arises because otherwise irrelevant cosine operator of Eq.~(\ref{SG}) becomes marginal at this point (as we show in the next section, one has $K_0=\frac{1}{2}$ at the peak value). The cosine term makes a contribution to the central charge $c=1+{\cal{O}}(g^3)$ and the pronounced peak value serves as an effective transition point for the finite-size system. In the thermodynamic limit the peak value $c_{peak} \rightarrow 1$ as $L\rightarrow \infty$, whereas $\kappa_{peak}\rightarrow \kappa_c$. As shown in Fig.~\ref{fig:C}(c), the scaling of data at the peak position does not follow $c=1+{\cal{O}}(1/L^2)$, which implies that there are usual logarithmic corrections $\propto  (-1)^l\sqrt{\ln{l}}/l$ for the TLL parameter in Eq.~(\ref{Fluctuation}) ~\cite{Nishimoto,Song2010}. We thus perform an accurate estimate of the critical parameter $\kappa_c$ from the calculation of the TLL parameter $K_0$, which we present below.
    We note that an abrupt decrease of the central charge from unity as a function of $\kappa>\kappa_c$ captures the transition from the TLL phase to the bond-order gapped phase, and hence the renormalization of $c$ to $0$.
    
    There is also an extended region of large values of $c$ for $-1<\kappa<-0.2$. This feature also comes from irrelevant contributions, not considered within the bosonized theory in the previous section. As one increases the system size, the central charge $c(\kappa)$ gets closer to $c=1$, so that the system renormalizes to the pure TLL.

\begin{figure}[t]
\includegraphics[width=\columnwidth]{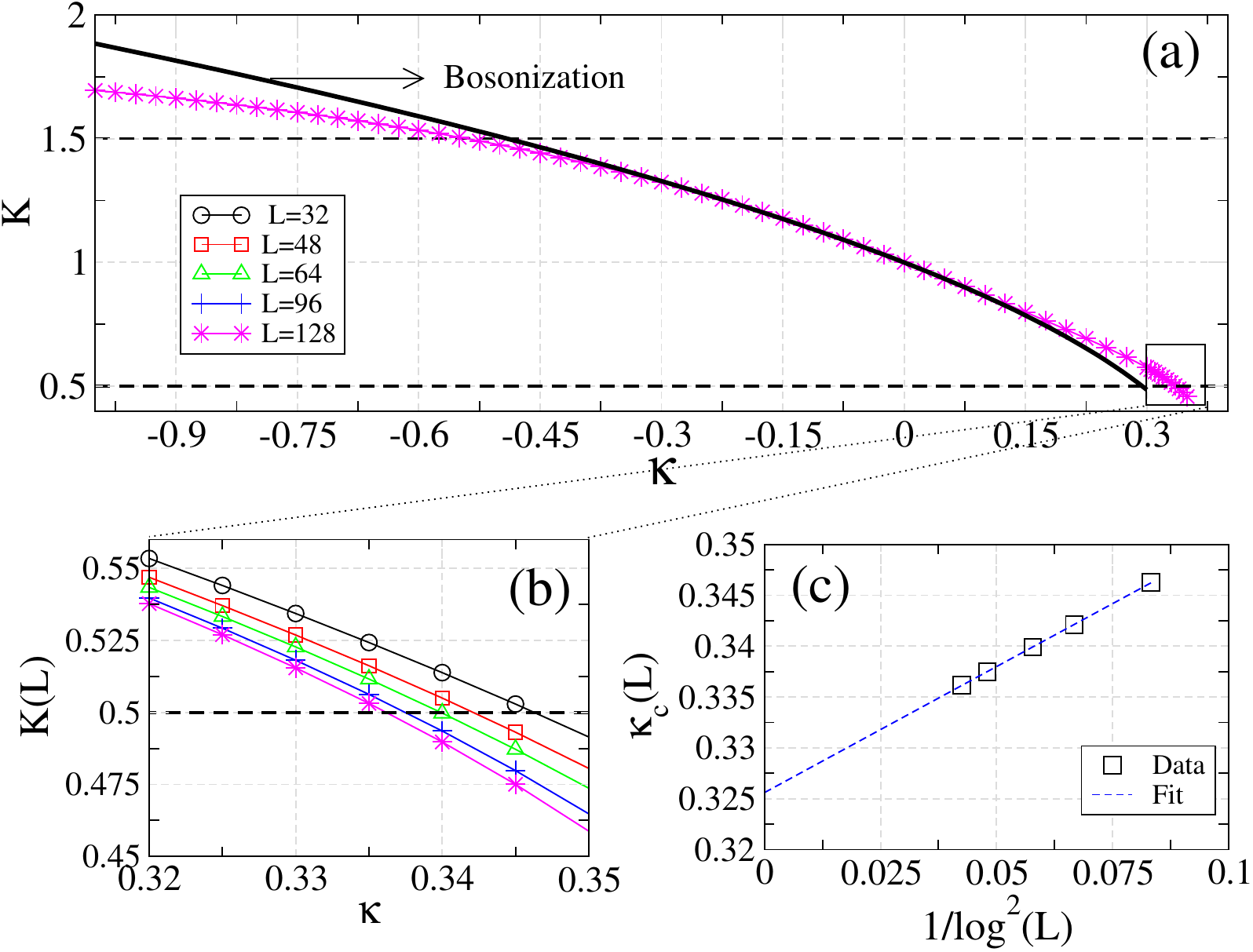}
\caption{ (a) DMRG result for the TLL parameter $K_0$ as a function of $\kappa$ for $L=128$ (symbols). The solid line represents the bosonization result of Eq.~(\ref{K_expression}) for $K_0$.  The results for the other considered $L$ around the critical point are plotted in (b). The finite-size critical data $K_0(L)$ obeys $1/\log^2(L)$ scaling as demonstrated in (c). }
\label{fig:K_clean}
\end{figure}

    {\it{ Clean TLL parameter $K_0$.}} The results of our calculations for the TLL parameter $K_0$ are presented in Fig.~\ref{fig:K_clean} (a). Remarkably, the analytical result of bosonization (solid line) and DMRG results (symbols $L=128$) are in agreement in the parameter range $|\kappa|<0.25$. For larger values of $|\kappa|$ the discrepancy between the two is large and grows with $\kappa$, which arises due to irrelevant terms excluded within our bosonization analysis.
    
        For positive $\kappa>0$ the TLL parameter has dominant bond-order-wave correlations and $K_0<1$. At $K_0<1/2$ the cosine term in Eq.~(\ref{SG}) becomes relevant and opens a gap in the spectrum. This is expected to occur at $\kappa\approx0.295$ from the bosonization result of Eq.~(\ref{K_expression}). To extract the numerical value, in Fig.~\ref{fig:K_clean}(b) we plot our numerical data around the expected critical point $K_0(L)$ for all considered system sizes. Since this phase transition belongs to the BKT universality class, in order to accurately locate the critical point in the thermodynamic limit we use the scaling argument $\kappa_c-\kappa_c(L)  \propto  \left[\log(L)\right]^{-2}$, where $\kappa_c$ is the critical parameter in the thermodynamic limit. The result of such a scaling is shown in Fig.~\ref{fig:K_clean}(c). Numerical data perfectly obey this scaling law and one extracts the value $\kappa_c=0.3256(2)$ from the fitting procedure. 
        
         For $\kappa<0$ the TLL parameter grows with $|\kappa|$ (from unity at $\kappa=0$) and one has enhanced SF correlations. In this regime, our results for $K_0$ show that $|K_0(L=96)-K_0(L=128)|\sim 10^{-4}$, and we accept the value $K_0(L=128)$ as the thermodynamic limit value. Remarkably, for $\kappa< 0.541$ (from the bosonization result one obtains $\kappa <- \frac{5\pi}{32} \approx -0.49$) the value $K_0>3/2$. One expects that at these values of $\kappa$ the disordered system keeps algebraic correlations at a finite disorder amplitude $h<h_c$, whereas for larger values $h$ the power-law decrease changes to an exponential one, which is the feature of the BG phase. As we show in the next section, this is indeed the case.

\section{Numerical results: disordered case}
\label{ResDis}

\begin{figure}[t]
\includegraphics[width=\columnwidth]{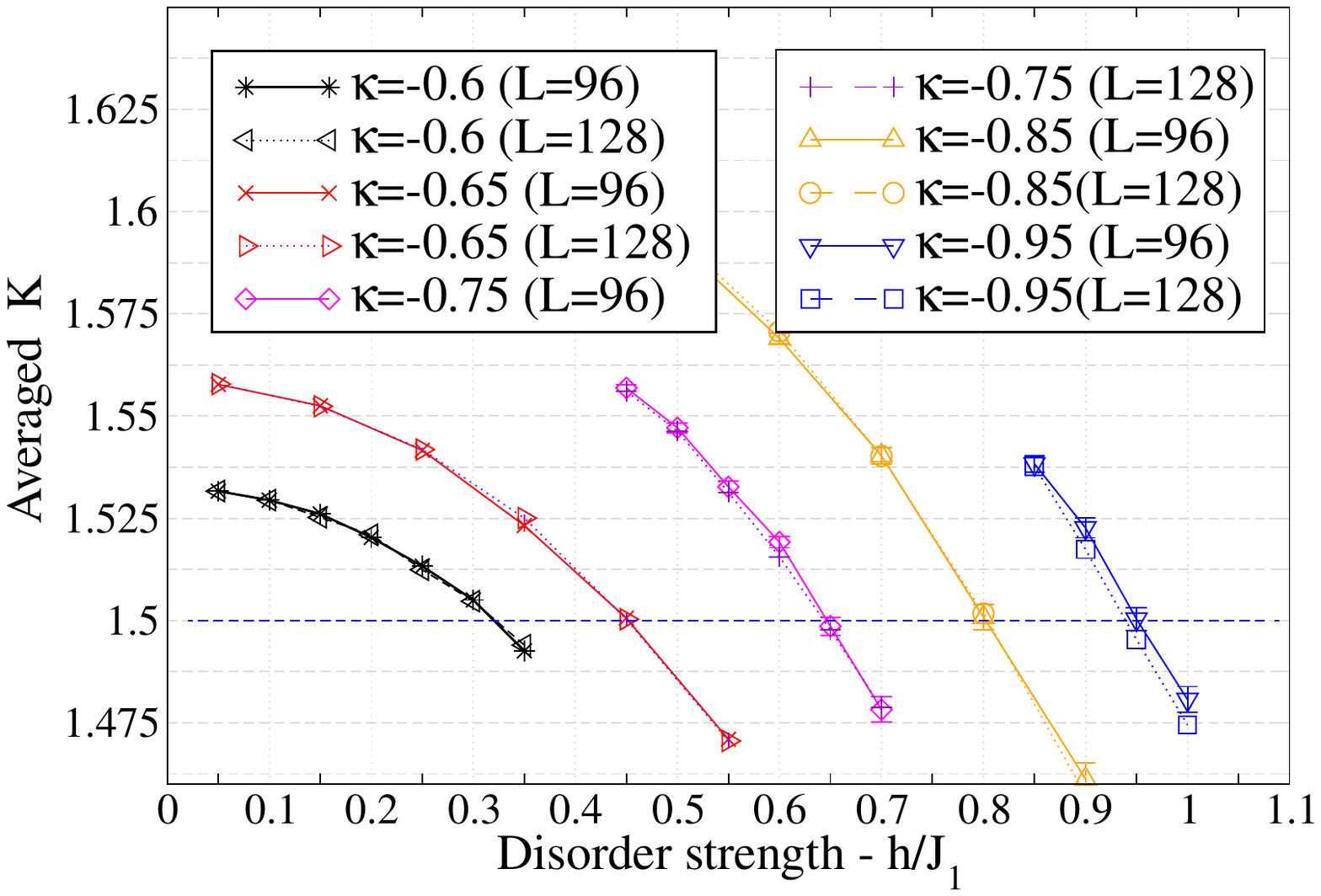}
\caption{Disorder-averaged values of the TLL parameter $K$ as functions of the disorder strength $h/J_1$ for several values of the parameter $\kappa$. The critical point is determined from the crossing of the $K(h)$ curves with the value $K_c=3/2$. The considered system sizes are $L=96$ and $L=128$. }
\label{fig:K_dis}
\end{figure}
 
We first assume that the BKT transition at finite disorder strength $h_c(\kappa)$ for $\kappa < -0.541$ obeys the GS scenario, i.e. occurs at $K_c=3/2$. For locating the critical line in the $\kappa-h$ space, we chose several values of $\kappa$ and calculated the $h$-dependent TLL parameter $K(h)$. The results of the calculation for $L=96$ and $L=128$ are presented in Fig.~\ref{fig:K_dis}. At finite disorder $h<h_c$ the TLL parameter $K(h)<K_0(h=0)$ decreases from its clean value and crosses the critical line $K_c=3/2$ at $h=h_c$. For the smallest considered $\kappa=-0.6$ the critical field $h_c/J_1 \approx 0.32$. As expected, for larger values of $\kappa$ one needs larger values of the disorder strength $h_c$, reaching $h_c/J_1\approx 0.95$ at $\kappa=-0.95$. As one can see from the plot, our results for the critical fields for $L=128$ coincide with the ones for $L=96$ within the considered accuracy. We also checked the consistency of these results using open boundary condition for the larger system size $L=384$ (not shown, see Ref.~\cite{Doggen2017} for details of similar calculations).
    We now justify the assumption that the considered BKT transition indeed obeys the GS scenario. For this, we provide arguments based on the following calculations:

\begin{figure}[t]
\includegraphics[width=\columnwidth]{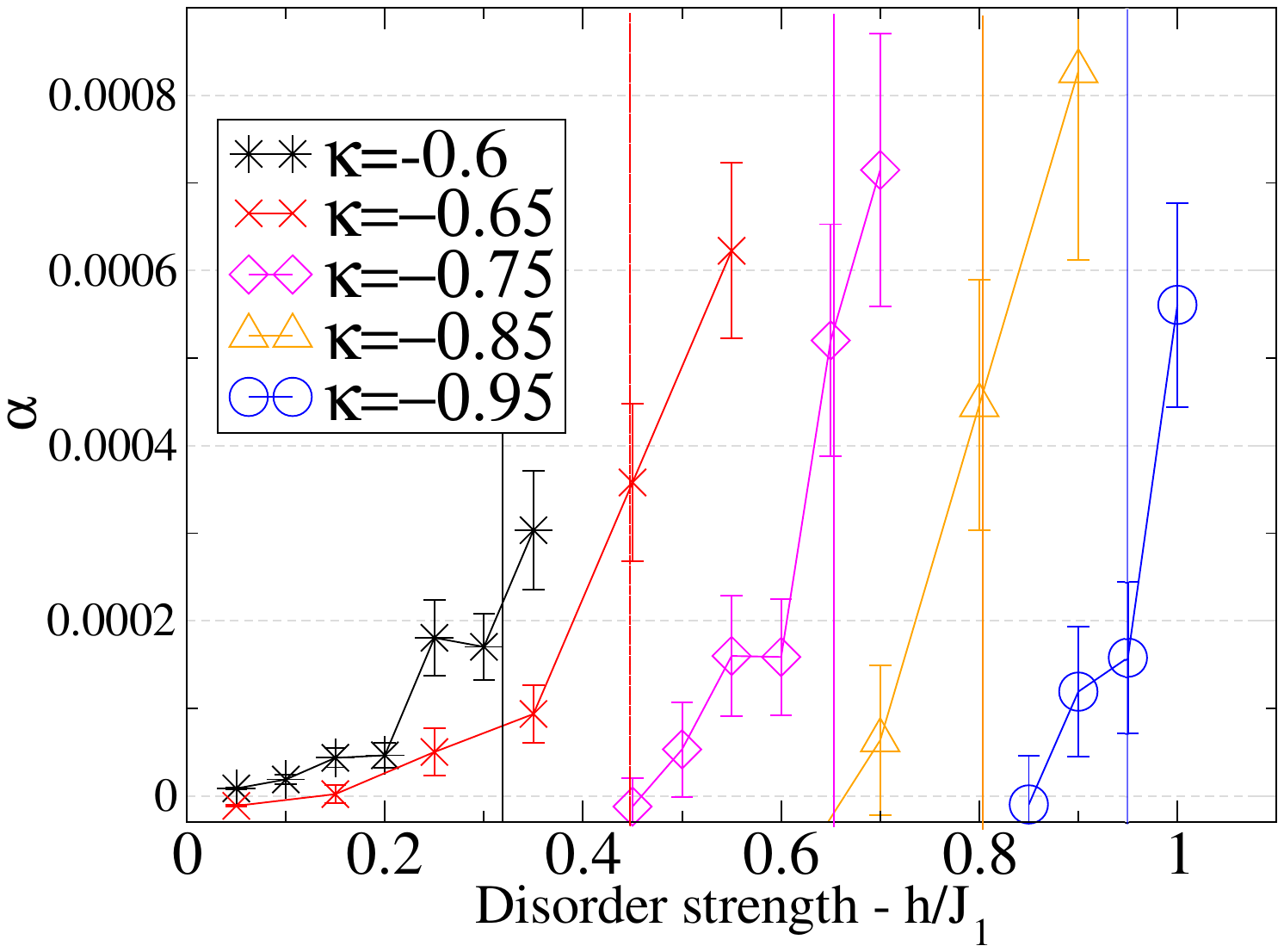}
\caption{ Numerical values of $\alpha$ obtained using the fitting procedure as a function of the disorder strength $h/J_1$. Vertical lines represent the values of $h_c/J_1$ obtained from the criterion $K(h_c)=3/2$. The results are presented for $L=128$. }
\label{fig:Alpha}
\end{figure}

{ \it Calculation of $\alpha$} - Our confirmation of the GS scenario is based on the  behavior of the disorder-averaged correlation function at long distances. In the SF phase this quantity decays {\it algebraically} with $\alpha=0$, whereas in the BG phase {\it exponential} decay with $\alpha>0$ is expected. Our results for $\alpha$ at several values of $\kappa$ are presented in Fig.~\ref{fig:Alpha}. For all considered values of $\kappa$, an increase of $\alpha$ with $h$ is sharp and it occurs in the vicinity of $h_c$ determined from the calculation of $K$ (vertical lines in the plot). These results imply that the transition indeed occurs in the vicinity of $K\approx3/2$. 
    
\begin{figure}[t]
\includegraphics[width=\columnwidth]{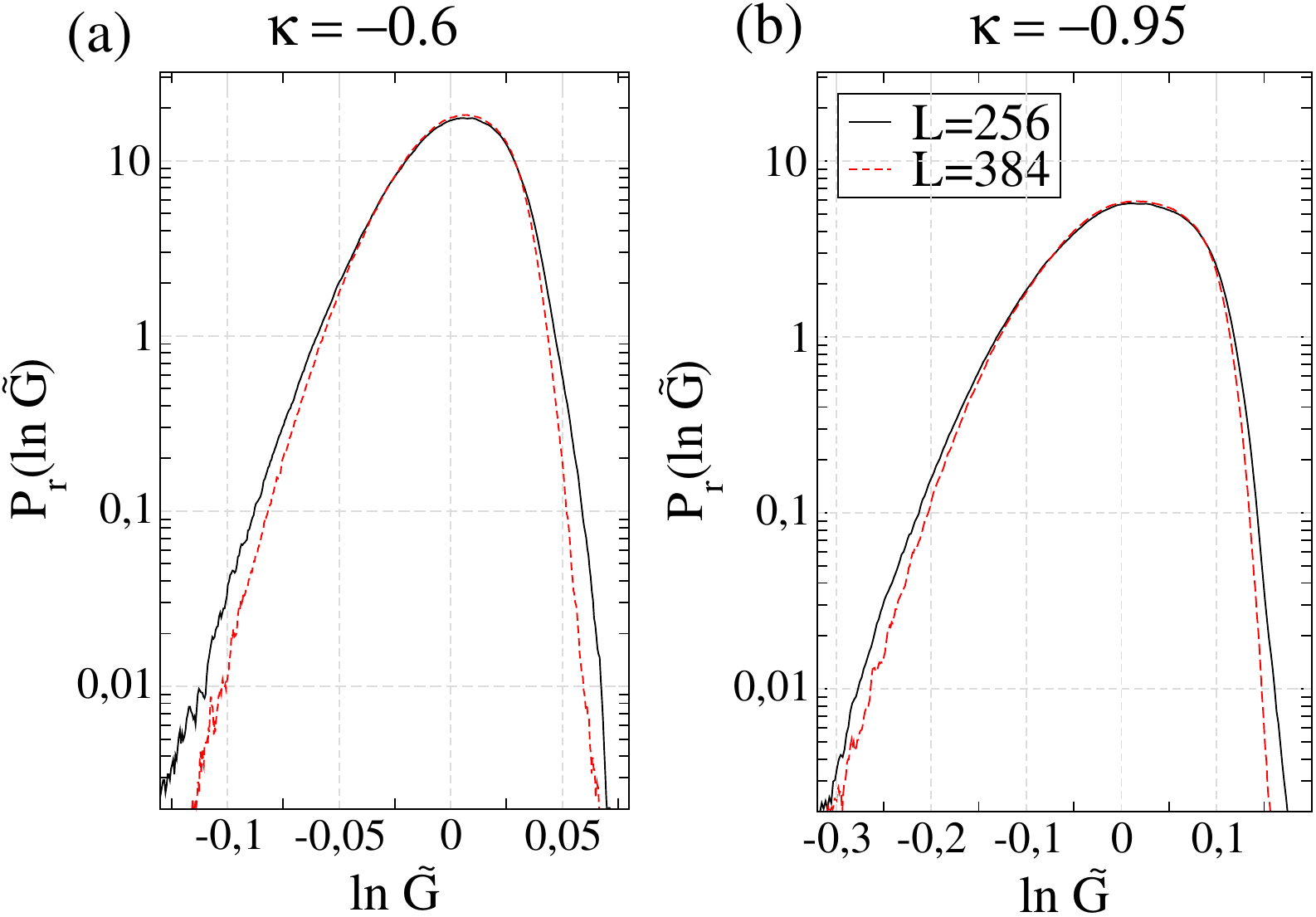}
\caption{ Fluctuation distribution of the correlator $\ln \Tilde{G}=\ln G- \overline{\ln G}$ (fixed $r\in \left[30:100 \right]$) at the critical disorder $h_c$ for $\kappa=-0.6$  ($h_c\approx0.32$) and $\kappa=-0.95$ ($h_c\approx0.94$). Distributions show self-averaging behavior and no weak-link tails are observed.}
\label{fig:G}
\end{figure}

{ \it Critical distributions of $G$} - To finalize our arguments, we followed Ref.~\cite{Doggen2017} and calculated the distribution of the fluctuation of the correlator $\ln \Tilde{G}=\ln G - \overline{\ln G}$ ($\overline{\cdot \cdot \cdot}$ denotes disorder-averaging) at distances $r\in \left[30:100\right]$ for the critical disorder and the system sizes $L=256$ and $L=384$ with open boundaries. In order to avoid finite-size effects, we included in our analysis only $70\%$ of lattice sites from the middle of the chain and excluded the remained edge sites. We considered $\kappa=-0.6$ and $\kappa = -0.95$ with $h_c/J_1\approx 0.32 $ and $h_c/J_1 \approx -0.94$, respectively. The results are presented in Fig.~(\ref{fig:G}). For both values of $\kappa$ one observes self-averaging behavior: when the system size is increased from $L=256$ to $L=384$, both distributions shrink. This is in sharp contrast with the weak-link scenario, where a {\it{self-similar}} behavior of the fluctuation distribution was observed~\cite{Doggen2017}. More importantly, the fluctuation distributions for both $\kappa$ do not posses exponential tails. The latter served as a smoking gun for the weak-link scenario, reported in Ref.~\cite{Doggen2017}. All these features support our assumption of the GS scenario and the absence of weak-link physics in our model for all values of $\kappa$.

The solely exhibited GS scenario in our model is in sharp contrast with the case of disordered 1D XXZ model. In the latter model, the weak-link scenario was numerically demonstrated when the critical disorder was larger than the bandwidth (when the Ising anisotropy $ \Delta>-1$)~\cite{Doggen2017}. The absence of the weak-link scenario in our model can be qualitatively explained as follows. In the case of the 1D XXZ model in the vicinity of the ferromagnetic phase transition one has $\Delta>-1$ and the clean TLL parameter $K_0$ behaves as $\propto \frac{1}{\sqrt{1+\Delta}}$. This shows that in this regime the model can be mapped onto the model of weakly interacting Bose gas in the low-energy limit, since for the free Bose gas one has $K_0=\infty$. Weak short-range interaction between the bosons renormalizes the TLL parameter to $K_0<\infty$ and introducing disorder in the form of a random potential one achieves the model studied in Refs.~\cite{Altman1,Altman2,Altman3,Boris}. At strong disorder, bosons form localized clusters within the Lifshitz tails. Since the density of states in the tail is exponentially small, these clusters are well separated. They do not overlap with each other and one has the insulating BG phase. When the disorder strength is comparable to the hopping amplitude, these clusters merge and form the SF phase. In the spin-1/2 notation, the clusters correspond to the domains of parallel spins formed in the vicinity of the ferromagnetic transition.
We also note that there exists exact mapping between the 1D XXZ model in the vicinity of ferromagnetic phase transition and the integrable Lieb-Liniger gas of bosons, obtained via the Bethe Ansatz method~\cite{LiebLiniger1,LiebLiniger2}. 

\begin{figure}[t]
\includegraphics[width=\columnwidth]{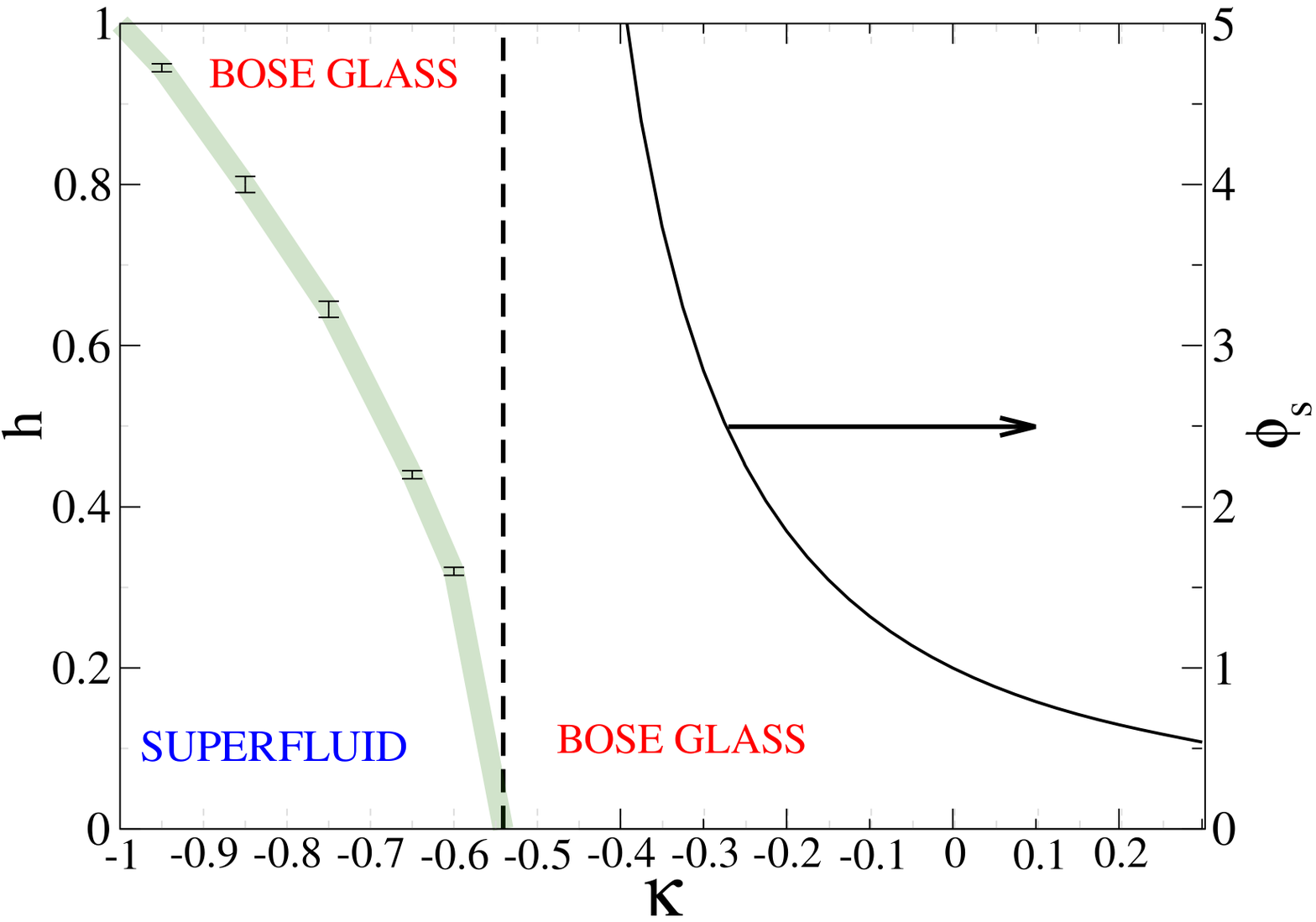}
\caption{Phase diagram of the model in the $h-\kappa$ plane and the exponent $\phi$ as function of $\kappa$ (solid curve). The dashed vertical line at $\kappa \approx -0.54$ separates the exhibited two regimes: for $-1\leq\kappa \lesssim -0.54$ and $h<h_c$  the SF phase persists. At larger disorder strength $h>h_c$ one enters the BG phase. In the second regime for $\kappa>-0.54$ arbitrarily weak disorder drives the system into the BG phase. The solid strip serves as eye-guide.}
\label{fig:Phase}
\end{figure}

    In the model of our study, the clean TLL parameters in the regime of dominant SF correlations are $K_0\sim 3/2$ and the mapping onto the weakly interacting Bose gas is violated. Instead, dual bosons strongly interact and become localized at finite disorder via the GS scenario, and the weak-link scenario is not exhibited. The phase diagram of our model is presented in Fig.~\ref{fig:Phase}.

\section{Conclusions}
\label{Conc}

In this work, we provided numerical evidence for the TLL - BG phase transition in the 1D fermionic system with pair hoppings. For sufficiently large pair hopping amplitudes, SF correlations in the bulk of the liquid are strongly enhanced with the clean TLL parameter $K_0>3/2$. In this regime, weak on-site disorder is an irrelevant perturbation and the TLL phase with algebraically decaying correlations persists at weak disorder. On the contrary, strong disorder drives the system to the BG phase via the BKT mechanism. We demonstrate that the transition follows the Giamarchi-Schulz scenario at $K_c=3/2$ in the thermodynamic limit. In the regime of weak pair hoppings with the clean TLL parameter $K_0<3/2$ arbitrarily weak disorder is relevant in the RG sense and 
one enters the BG phase at any finite disorder with the disorder-dependent correlation length $L_c=h^{-2\phi_s}$. The exponent depends on the clean TLL parameter as $\phi_s=(3-2K_0)^{-1}$. The form of the exponent in the localization regime together with the phase diagram of the model are presented in Fig.~\ref{fig:Phase}. 
\begin{acknowledgements}
 This research was supported in part through computational resources of the HPC facilities at HSE University~\cite{Kostenetskiy_2021}. MSB thanks Basic Research Program of HSE for the provided support. 

\end{acknowledgements}

\end{document}